\documentclass[%
 reprint,
superscriptaddress,
 amsmath,amssymb,
 aps,
showkeys
]{revtex4-2}

\usepackage{xcolor}
\usepackage[inkscapeformat=png]{svg}
\usepackage{hyperref}
\usepackage{upgreek}
\usepackage{siunitx}
\usepackage{subcaption}
\usepackage{graphicx}
\graphicspath{{Main/}}
\usepackage[thinc]{esdiff}
\usepackage{dcolumn}
\usepackage{bm}

\usepackage{xcite}

\graphicspath{{figures/}{./}}

\usepackage{caption}
\usepackage{subcaption}
\usepackage{xr}
\makeatletter
\newcommand*{\addFileDependency}[1]{
  \typeout{(#1)}
  \@addtofilelist{#1}
  \IfFileExists{#1}{}{\typeout{No file #1.}}
}
\makeatother

\begin{document}

\preprint{APS/123-QED}

\title{Viscotaxis of Beating Flagella at Surfaces}

\author{Shubham Anand}
 \email{s.anand@fz-juelich.de}
 \affiliation{%
 Theoretical Physics of Living Matter, Institute for Advanced Simulation and Institute for Biological Information Processing, Forschungszentrum J\"{u}lich,  52425 J\"{u}lich, Germany
}%

\author{Jens Elgeti}
 \email{j.elgeti@fz-juelich.de}
\affiliation{%
 Theoretical Physics of Living Matter, Institute for Advanced Simulation and Institute for Biological Information Processing, Forschungszentrum J\"{u}lich,  52425 J\"{u}lich, Germany
}%

\author{Gerhard Gompper}%
 \email{g.gompper@fz-juelich.de}
\affiliation{%
 Theoretical Physics of Living Matter, Institute for Advanced Simulation and Institute for Biological Information Processing, Forschungszentrum J\"{u}lich,  52425 J\"{u}lich, Germany
}%

%

\date{\today}

\begin{abstract}
Many biological microorganisms and artificial microswimmers react to external cues of
environmental gradients by changing their swimming directions. We study here the behavior 
of eukarytic flagellated microswimmers in linear viscosity gradients.
We employ a model of flagellum consisting of a semi-flexible filament with a travelling 
wave of spontaneous curvature to study viscotaxis of actively beating flagella in two
spatial dimensions.  The propulsion of the flagellum in a fluid due to a 
hydrodynamic friction anisotropy is described by resistive-force theory.
Using numerical simulations and analytical theory, we show that beating flagella 
exhibit positive viscotaxis, reorienting themselves toward higher viscosity areas. 
We quantify this behavior by characterization of the dependence of the rotational velocity 
on beat amplitude, swimming speed, and wave length. We also examining the effects 
of asymmetric flagellar wave forms, which imply circular trajectories in the absence of 
viscosity gradients; here, large asymmetry leads to trochoid-like trajectories 
perpendicular to the gradient in the form drifting circles. The flagellar deformability 
strongly reduce the beat amplitude and the viscotatic response. The viscotatic response 
is shown to be captured by a universal function of the sperm number. 

\end{abstract}

\maketitle


\section{Introduction}
\label{sec:introduction}

Sperm cells move in complex environments to find and reach the egg, such as the female reproductive 
tract for species with internal fertilization, or open water for species with external fertilization, 
where both eggs and sperm are released into the bulk fluid.
This process poses several key challenges for sperm motion: How do sperm steer? How do they change 
their swimming direction? How do sperm sense and react to different environmental cues? 
How do environmental patterns such as flow or viscosity gradients influence their trajectory?
These are important issues to be clarified in order to understand how sperm can navigate their 
tortuous journey towards the egg \cite{alvarez_2014_review, miki2013sperm_rheo}. 

For an active steering response, two mechanisms have been proposed for steering \cite{gong2020steering}: 
An average curvature of the flagellum, which leads to a curved or helical trajectory 
\cite{elgeti2010hydrodynamics, friedrich2010high}, and higher harmonic beat frequencies  
\cite{saggiorato2017human}. Also, buckling of the flagellum under the load of the propulsion force
of the flagellum has been suggested \cite{smith2009bend}.
This steering response can be purposefully employed by the cell for spatial orientation, 
for example in chemical gradients \cite{yoshida2013species, jikeli_2015_natcommun, strunker2015physical, hong2007chemotaxis}. 
This chemotactic behavior is typically controlled by biochemical processes in the cell. 

In general, the movement of biological microswimmers, such as sperm, can also be significantly influenced 
by changes in the physical properties of their environment \cite{elgeti2015physics}, 
where microswimmers can reorient in external gradients by physical forces exerted on their
bodies. 
This includes phenomena like rheotaxis \cite{zoettl_2012_rheo, marcos_2012_rheotaxis, qi_2020_rheotaxis}, 
where microswimmers can swim upstream in flow through microchannels, and gravitaxis 
\cite{tenhagen_2014_gravitaxis}, where they move against the gravitational field. 

We focus here on the much less explored effect of viscotaxis,
the motion of microswimmers in viscosity gradients \cite{liebchen2018viscotaxis,shaik2021hydrodynamics}. 
Different kinds of microswimmers have been found to respond to the viscosity 
gradients, depending on their shape or hydrodynamic propulsion mechanisms. Three-bead model microswimmers
have been shown to display positive viscotaxis due to a friction imbalance \cite{liebchen2018viscotaxis}
while squirmers display negative viscotaxis due to enhanced surface propulsion in high-viscosity
fluids  \cite{shaik2021hydrodynamics, datt2019active}. Previous studies have also considered the 
dynamics of microswimmers in response to sharp viscosity gradients, as it occurs at a soft, penetrable 
interface in a binary mixture of two fluids with different viscosities \cite{feng2022dynamics, gong2023active, coppola2021green}.   

Specifically, we investigate the movement of active eukaryotic flagella, with a snake-like,
beat pattern, in fluids with spatially varying viscosity.
Our combined simulation and theoretical modeling study describes a flagellum as a semi-flexible 
filament subject to a traveling bending wave \cite{yang2008cooperation}. 
Our explicit modelling of the beating flagellum allows detailed predictions of viscotaxis in terms of
the physical properties of the flagellum, such as bending rigidity of the flagellum, average spontaneous
curvature, beat frequency, and wave length. 

Here, it is interesting to note that the inherent spontaneous 
curvature of flagella is an important parameter in sperm navigation, 
sperm cells of various species exhibit a range of morphological shapes, which are often not 
symmetrical. This asymmetry of some sperm cells facilitates directional steering toward favorable 
environments, e.g. sea urchin sperm have been shown to utilize spontaneous curvature to actively 
adjust their movement in chemotactic response \cite{friedrich2007chemotaxis,friedrich2008stochastic,gong2020steering}.

Sperm often move along surfaces, as their propulsion and hydrodynamic interactions imply an effective,
dynamic attraction to surfaces \cite{berke_2008_swimsurface, elgeti_2009_swimsurface, elgeti_2016_review}. 
We model this swimming-near-surfaces behavior by considering flagellar motion in two spatial dimensions. 
Our simulations reveal that individual, symmetrically beating flagella generally
exhibit positive viscotaxis, i.e. they reorient to align with a viscosity
gradient, moving toward regions with higher viscosity. However, asymmetrically beating
flagella, with average spontaneous curvature, can swim at finite angles with the gradient or move in 
drifting circles perpendicular to the gradient direction. These results are valuable 
for controlling sperm motion in microfluidic devices \cite{kantsler2013ciliary, denissenko2012human, nosrati2014corner,nosrati2016predominance,rode2019sperm}.

\section{Model and Methods}
\label{sec:model}

\begin{figure}
     \centering
     \begin{subfigure}[b]{0.235\textwidth}
         \centering
         \includegraphics[width=\textwidth]{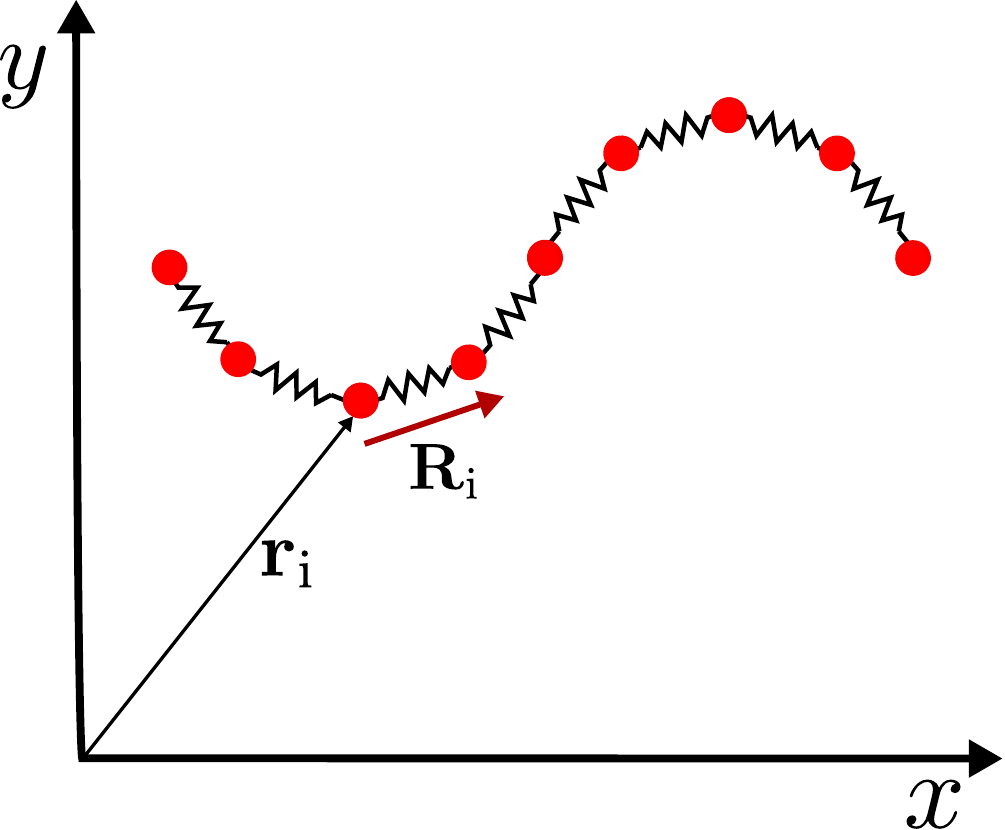}
     \end{subfigure}
     \hfill
     \begin{subfigure}[b]{0.235\textwidth}
         \centering
         \includegraphics[width=\textwidth]{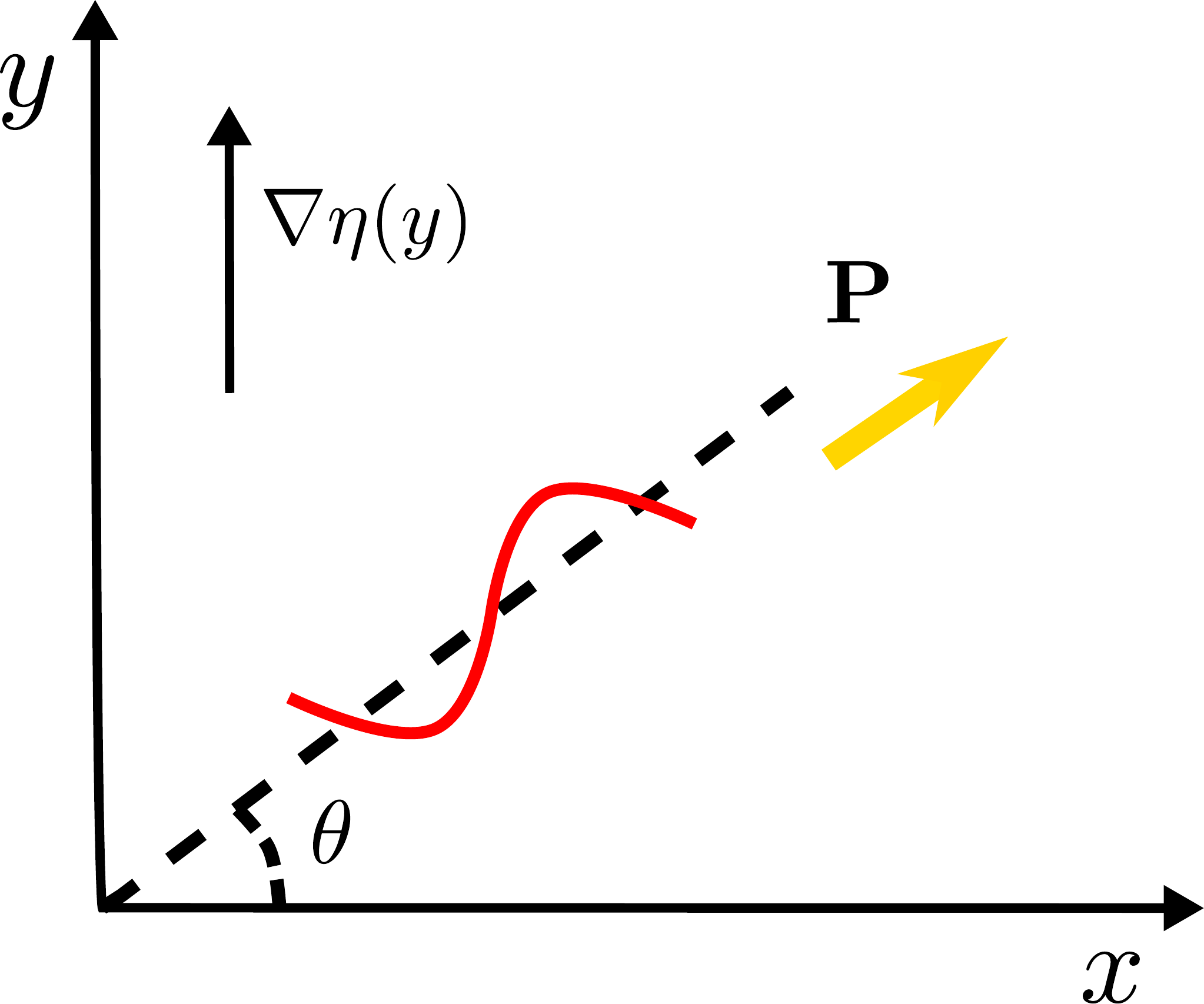}
     \end{subfigure}
\caption{\textbf{Model and swimming geometry in a viscosity gradient.} (a) Graphical representation of the 
    model for actively beating flagellum, where 
    $\mathbf{r}_{\mathrm{i}}$ is the position vector of particles and $\mathbf{R}_{\mathrm{i}}$ 
    is the bond vector. (b) Schematic of the swimming geometry in a viscosity gradient, with 
    polarity vector $\bf{P}$, viscosity-gradient direction ${\bf e}_y$, and orientation angle $\theta$, 
    where $\sin(\theta)={\bf P} \cdot {\bf e}_y$.}
    \label{fig:model}
\end{figure}

The sperm flagellum is modelled as an actively beating semi-flexible filament in two dimensions,
constructed of $N$ beads connected by bonds to form a linear 
chain, see Fig.~\ref{fig:model}(a). The structure and dynamics of the 
flagellum is described by the position vectors $\mathbf{r}_{\mathrm{i}}$ and velocity vectors 
$\mathbf{v}_{\mathrm{i}}$ of each bead. The total force 
$\mathbf{f}_{\mathrm{i}}$ acting on each bead includes the bond force, a curvature-elastic
force, 
and a viscous force due to friction with the embedding medium. The dynamics of 
each bead within the flagellum is then governed by the equation of motion
\begin{align}
    \frac{d\mathbf{r}_{\mathrm{i}}}{dt} &= \mathbf{v}_{\mathrm{i}} \ \ , \ \ 
    \frac{d\mathbf{v}_{\mathrm{i}}}{dt} = \frac{\mathbf{f}_{\mathrm{i}}}{m} \, .
    \label{eq:model}
\end{align}
Here, a small mass $m$ is used for computational efficiency.
Effectively, using a finite mass and velocity allow us to use Velocity-Verlet integration, corresponding 
to using a higher order integrator with a physically interpretable control parameter, the 
mass $m$ 
\cite{isele2016dynamics}. However, care has to be taken to reproduce overdamped dynamics, which is 
achieved by setting mass $m$ and friction coefficient $\xi$ such that the viscous relaxation time $m/\xi$ is much 
shorter than all other relevant timescales, in particular the beat period $\tau$. 

A constant separation between neighboring beads is determined by a harmonic bond potential, 
\begin{align}
    U_{bond} = \frac{k}{2}\sum_{\mathrm{i}=0}^{N-1}\left(\left|\mathbf{r}_{\mathrm{i+1}} - \mathbf{r}_{\mathrm{i}}\right| - \mathrm{b}\right)^{2} \, ,
\end{align}
where $\mathrm{b}$ is the rest length of the spring. To impose a smooth flagellar shape during the 
beating motion, and a propagating wave in the beat pattern, we employ the bending potential 
\begin{equation}
    U_{bend} = \frac{\kappa}{2\mathrm{b^{3}}}\sum_{\mathrm{i}=0}^{n-3}
          \left(\mathbf{R}_{\mathrm{i+1}} - \Re(bC_{\textrm{flag}})\mathbf{R}_{\mathrm{i}}\right)^2
    \label{eq:curvature_Hamilonian}
\end{equation}
where $\mathbf{R}_{\mathrm{i}}$ represents the bond vector that connects two 
neighbouring monomers, with 
$\mathbf{R}_{\mathrm{i}} = \mathbf{r}_{\mathrm{i+1}} - \mathbf{r}_{\mathrm{i}}$, 
and $\kappa$ is the bending rigidity.

The local spontaneous curvature $C_{\rm{flag}}$ is incorporated by the rotation matrix $\Re(bC_{\textrm{flag}})$ in 
Eq.~\eqref{eq:curvature_Hamilonian}, which rotates a vector anti-clockwise by an angle of $bC_{\rm{flag}}$.
A time-dependent local spontaneous curvature \cite{yang2008cooperation}
\begin{align}
    C_{\rm{flag}}(s,t) = C_{0} + A_{c} \sin(-\omega t + qs)
    \label{eq:curv}
\end{align}
then creates a propagating bending wave along the flagellum, where $t$ is time, and $s$ the position 
along the flagellum, $\omega=2\pi/\tau$ (with beat period $\tau$) is the beat frequency, and 
$q = 2\pi/\lambda$ the beat wave number (with wave length $\lambda$). Equation \eqref{eq:curv} 
also contains is the curvature amplitude $A_{c}$, and a constant average spontaneous curvature 
$C_{0}$ of the flagellum.

In previous studies, it was observed that the beating pattern of bull sperm 
\cite{riedel2007molecular} and of human sperm \cite{saggiorato2017human} can 
be accurately represented by a single beating mode. Thus, the single mode in Eq.~\eqref{eq:curv} captures 
the essential characteristics of the beat motion.

The bending wave propagates along the flagellum, creating backward fluid motion
in the embedding viscous medium, and thereby simultaneously propelling the 
flagellum forward.
We do not model the fluid explicitly in our simulations, but employ resistive force theory (RFT) instead. Here, 
propulsion is facilitated by the fact that a slender object in a viscous fluid 
with viscosity $\eta$ experiences unequal friction coefficients $\xi_{\parallel}$ and $\xi_{\perp}$ 
(both depending linearly on $\eta$) for motion tangential and perpendicular to its contour. 
In the limit of large aspect ratio, the friction anisotropy of a stiff filament is $\xi_{\perp}/\xi_{\parallel}= 2$ 
\cite{brennen1977fluid, gray1955propulsion, lighthill1976j, yu2006experimental, lauga2009hydrodynamics, johnson1979flagellar}. 
Experimental data for bull sperm \cite{friedrich2010high} yields a somewhat smaller value of 
$\xi_{\perp}/\xi_{\parallel}= 1.81$. We apply resistive force theory to each bead in our chain, where 
the force depends on the local tangent and normal directions of the flagellum, with
\begin{align}
    \mathbf{f}_{\mathrm{i}}^{\mathrm{r}} = -\left(\xi_{\perp}\mathbf{\hat{n}}_{\mathrm{i}}\mathbf{\hat{n}}_{\mathrm{i}}^{T} 
    + \xi_{\parallel}\mathbf{\hat{t}}_{\mathrm{i}}\mathbf{\hat{t}}_{\mathrm{i}}^{T}\right)\boldsymbol{v}_{\mathrm{i}}, 
    \,\,\,  \xi_{\perp}, \ \xi_{\parallel} \sim \eta 
    \label{eq:frictionforce}
\end{align}

We employ flagellum length $L=Nb$ and beat period $\tau$ as the basic length and time scales in the simulations, respectively. 
To ensure a realistic representation of sperm flagella, we select parameters that imitate the shape 
of the natural flagellum beat. 
A large spring constant $kb^2$ is employed to 
keep the bond length nearly constant.
To ensure that the dominant deformation is bending rather than stretching, we choose $\kappa/L <<  kb^2$, typically $\kappa/L= 3.18\times 10^{-4} \, kb^2$. 
A curvature amplitude $A_{c}L = 6.28$
generates beat amplitudes $A_b$ of about 15\% of the flagellum length $L$. For more details
about the parameter selection in the simulations, see the Supporting Information (SI) \cite{supp}.
Using these parameters, symmetrically beating flagella propel with constant velocity, as expected.

For small enough beat amplitudes, the velocity is well described by \cite{gray1955propulsion},
\begin{equation} 
v_{flag} = -\frac{1}{2}\left(\frac{\xi_{\perp}}{\xi_{\parallel}}-1\right)A_{b}^{2}\omega q
\label{eq:GH}
\end{equation}
for the beat pattern $y(x,t)=A_{b}\sin(-\omega t + qx)$, where $A_{b}$ is the beat amplitude 
(see SI for details \cite{supp}). In addition, $A_{c}$ and $A_{b}$ are 
related by $A_{b}=\lambda^{2}A_{c}/4\pi^{2}$ for small beat amplitudes.

\section{Flagellar Motion in Viscosity Gradients}
\label{sec:sph}

The main focus of our study is the dynamics of flagellar motion within spatially 
varying viscosity fields. 
We model a slowly varying viscosity as a simple linear gradient in viscosity gradient,
\begin{align}
    \eta(y) = \eta_{0}(1 + \alpha (y-y_{cm})/L),\ \ \ \ \xi_{\perp},\xi_{\parallel}\sim \eta(y) \, ,
    \label{eq:etay}
\end{align}
where $y_{cm} = \sum_{i} m_{i}y_{i}/\sum_{i} m_{i}$ represents the 
$y$-coordinate of the center of mass. Here, $\alpha$ is the (dimensionless) viscosity gradient coefficient 
and $\alpha < 1$. The spatial variation in viscosity is small on the scale of flagellum length $L$.

\subsection{Symmetric Flagellar Beat}
\label{sec:phasessph}

\subsubsection{Dynamics of Flagellar Orientation}
\label{sec:Viscosity_Gradients}

\begin{figure}
    \centering
    \includegraphics[width=0.45 \textwidth]{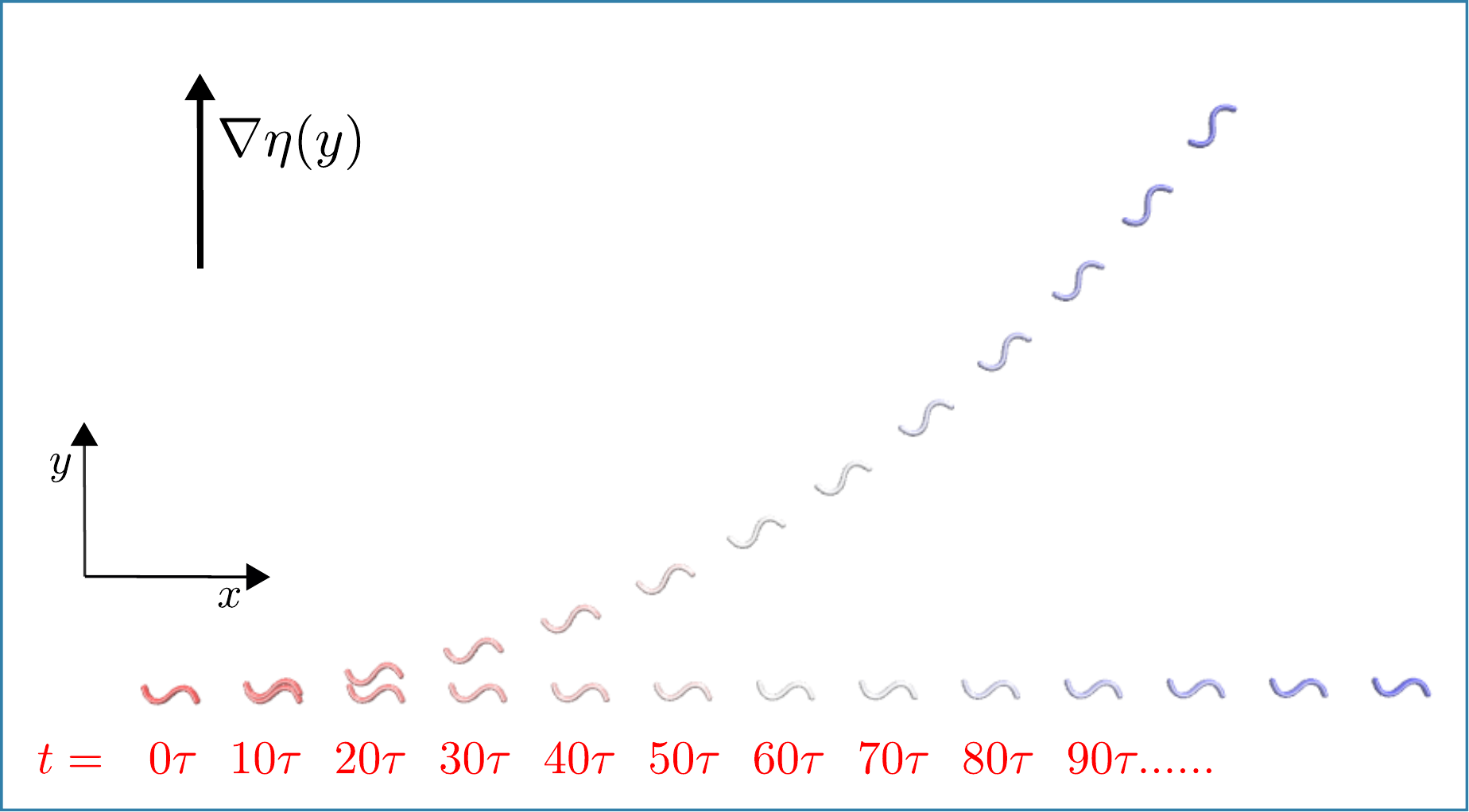}
    \caption{\textbf{Flagellum motion in viscosity gradient.} Superimposed snapshots of a symmetric beating flagellum under stroboscopic illumination 
demonstrating the progressive motion of the flagellum in viscosity gradient, with $\alpha=0.4$, 
$C_{0} = 0$ and $A_{c}L= 6.28$ and compared it's motion with the one without any gradient($\alpha=0.0$). 
Snapshot are separated by 10 beat periods $\tau$.}
\label{fig:etay}
\end{figure}

To gain insights into the behaviour of a beating flagellum in viscosity
gradients, we first consider a flagellum with a symmetric beat, 
i.e. without an average spontaneous curvature, $C(0)=0$ in Eq.~\eqref{eq:curv}.
The exemplary simulation trajectories in Fig.~\ref{fig:etay} show that the flagellum
adapts its motion to the viscosity gradient and migrates toward regions of higher viscosity.
This is denoted as {\em positive} viscotaxis,  see movie M1 \cite{supp}.
This migration is achieved by reorientation with rotational velocity $\Omega$ 
due to the presence of the gradient. 

\begin{figure*}[!htb]
     \centering
     \begin{subfigure}{0.31\textwidth}
         \centering
         \includegraphics[height=0.72\textwidth]{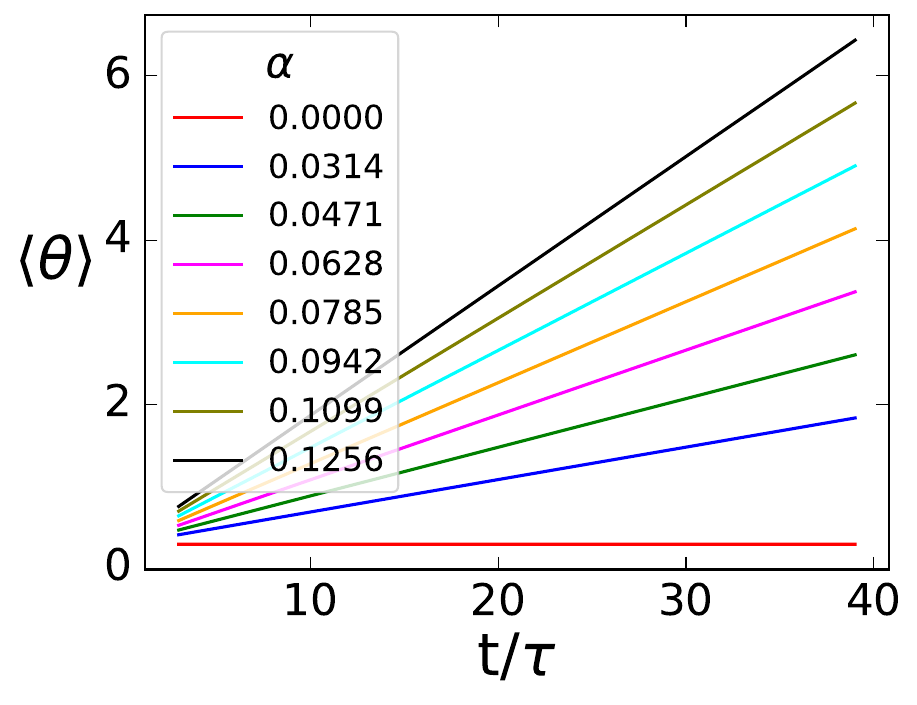}  
     \end{subfigure}
     \begin{subfigure}{0.31\textwidth}
         \centering
         \includegraphics[height=0.72\textwidth]{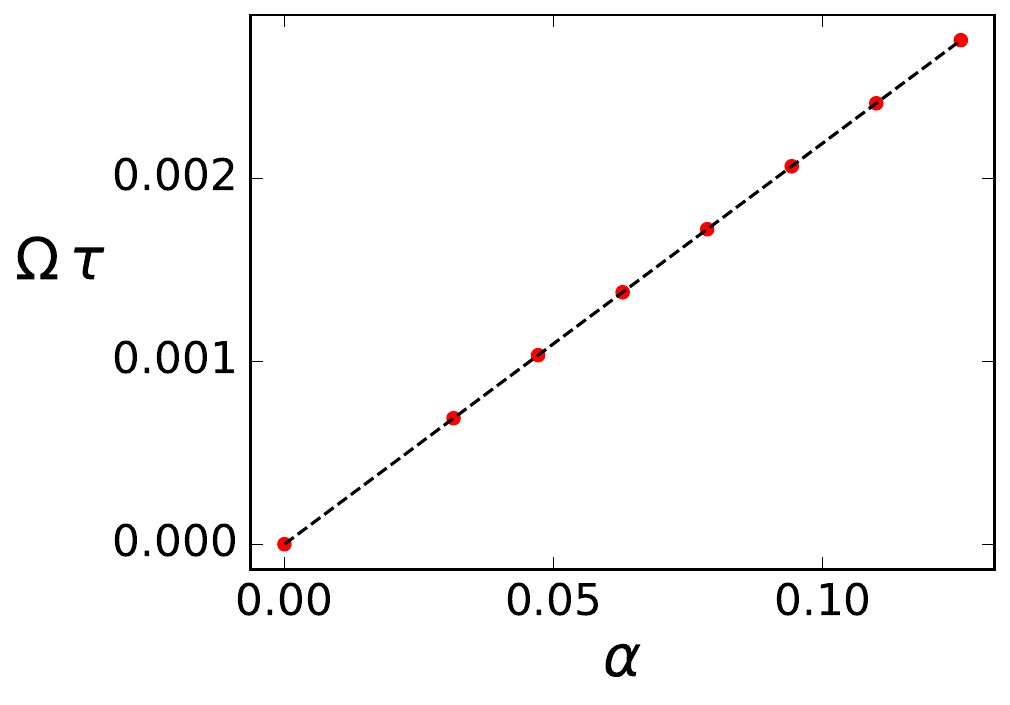}  
     \end{subfigure}
     \hspace{0.1cm}
     \begin{subfigure}{0.31\textwidth}
         \centering
         \includegraphics[height=0.72\textwidth]{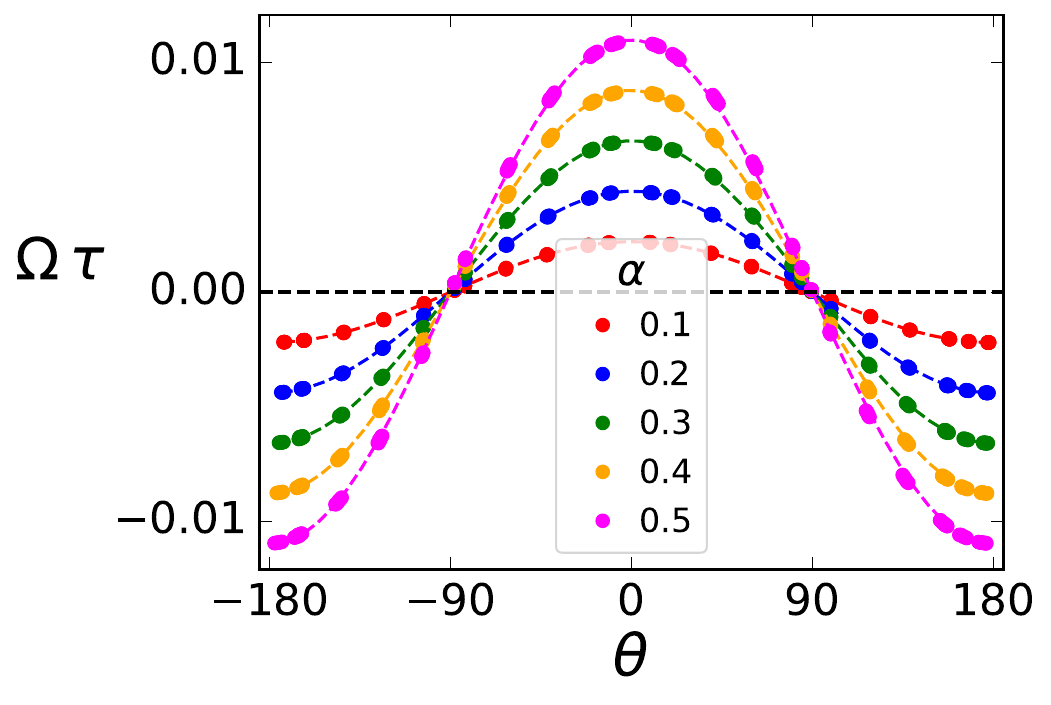}  
     \end{subfigure}
\caption{\textbf{Quantifying visotactic behavior.}
    (a) Averaged orientation of flagellum per beat $\left<\theta\right>$ for different viscosity 
    gradients $\alpha$ as a function of time.
    (b) Effect of different viscosity gradients on rotational velocity $\Omega$. Data points are 
    obtained from linear fits in (a). The dashed line
    is linear fit to Eq.~\eqref{eq:omegaofalpha}, yielding $\Omega_1$. In (a-b), the initial orientation 
    is chosen to be $\theta \approx 0$.
    (c) Rotational velocity $\Omega(\theta)$ as a function of orientation angle $\theta$ for different 
    viscosity 
    gradients $\alpha$. Solid points are the simulation data, dashed lines are fits to 
    Eq.~\eqref{eq:omegavstheta}. The single fit parameter $\Omega_1 \tau= 0.02196$ is the viscotactic 
    response coefficient. Error bars are smaller than the symbol size.}
    \label{fig:measure}
\end{figure*}

To quantify the viscotactic behavior, we study the time dependence of the orientation angle 
$\theta(t)$, which is the angle between the polarity vector (swimming direction) $\bf{P}$ 
and the positive $x$-axis, as shown in Fig.~\ref{fig:model}(b). 
This orientation angle (averaged over the periodic oscillations due to beating) increases 
roughly linearly in time, see Fig.~\ref{fig:measure}(a) -- until it saturates at $\theta=\pi/2$ 
for very long times.
Thus, we define the rate of reorientation (or angular velocity) $\Omega(\theta, \alpha)=d\theta/dt$.
The orientation rate increases with increasing viscosity gradient $\alpha$, as expected.
Expanding $\Omega(\theta, \alpha)$ as a power series for small $\alpha$, we get 
\begin{align} 
    \Omega(\theta,\alpha) = \Omega_{0} + \Omega_{1}(\theta)\alpha 
                       + \mathcal{O}(\alpha^{2})\approx\Omega_{1}\alpha \, .
    \label{eq:omegaofalpha}
\end{align}
Symmetry dictates that symmetric filaments ($C_0=0$) do not rotate without a gradient, and rotate 
in the opposite direction for opposite gradients. Thus $\Omega_{0} = 0$, and $\Omega_{1}(\theta)$ is 
the leading viscotactic coefficient. Figure~\ref{fig:measure}(b) shows that the rotational velocity 
$\Omega$ increases with increasing viscosity gradient $\alpha$.  
For the angular dependence, note that $\Omega(\theta)$ has to be $2\pi$ periodic. 
Mirror symmetry to the gradient (sperm facing "left" or "right" of the gradient both rotate towards the 
gradient) also means that $\Omega(\theta)=-\Omega(\pi-\theta)$, thus motivating a cosine expansion. 
Again the zero-order term vanishes due to symmetry, and to lowest order we get, 
\begin{equation}
    \Omega(\theta, \alpha) = \Omega_1 \alpha \cos(\theta),
    \label{eq:omegavstheta}
\end{equation}
see Fig.~\ref{fig:measure}(c).
Thus, $\Omega_1$ is the key viscotactic response coefficient. It quantifies how strong and in which 
direction the sperm react to a viscosity gradient. For Fig.~\ref{fig:measure}(a-b), the initial 
orientation is chosen to correspond to the maximum of $\Omega(\theta)$ in Fig.~\ref{fig:measure}(c), 
such that $\theta \approx 0$ and rotational velocity is obtained from the Eq.~\eqref{eq:omegavstheta}.

\subsection{Viscotaxis Mechanism and Analytical Estimate of Reorientation Rate}
\label{sec:barrierssph}

\begin{figure}
    \centering
    \includegraphics[width=0.38\textwidth]{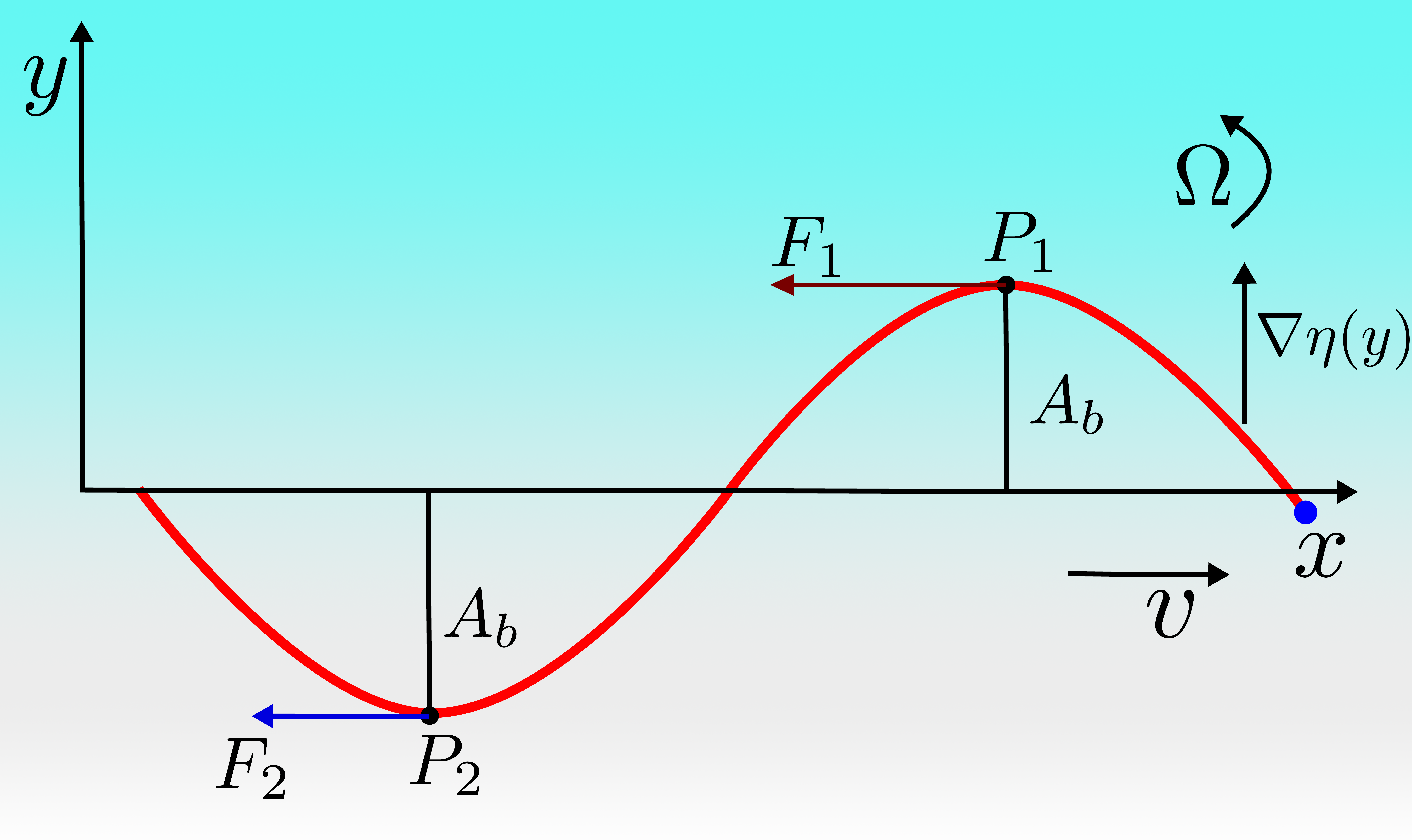}
    \caption{\textbf{Swimming in viscosity gradients.} Sketch of sperm swimming in viscous gradients. The viscosity gradient is in the $y$-direction, 
    while the flagellum swims in the $x$-direction and the blue small head represent the point from which the 
    propagation of the wave takes place. $P_1$ is further up the gradient than $P_2$, thus $F_1$ (longer maroon arrow) 
    is larger than $F_2$ (shorter blue arrow), resulting in a net torque toward the gradient.}
    \label{fig:diagram}
\end{figure}

We conjecture that the main underlying mechanism for reorientation in viscotaxis is the 
drag imbalance of the high- and low-viscosity sides 
of the swimming flagellum, as illustrated in Fig.~\ref{fig:diagram}. This imbalance in forces 
results in a net torque, causing the flagellum to rotate. To calculate this torque analytically, 
we imagine the beating filament as its sinusoidal shape traveling at constant velocity $v$ 
along the $x$-axis, perpendicular to the gradient direction ${\bf e}_y$.
The drag forces acting on the flagellum dependent on the local viscosity, and are dominated by 
$\xi_\parallel$ for small beating amplitudes. The torque around the center of mass then reads
\begin{align}
    T_A = \int_{-A_{b}}^{A_{b}} y \bar{\rho}(y) \, v \xi_{\parallel} \, \alpha y \, dy
    \label{eq:TA}
\end{align}
with the time averaged and $y$-projected mass density of the beating flagellum 
\begin{align} 
    \bar{\rho}(y) = \frac{1}{\tau}\int_0^\tau\int_{0}^{L}  \, 
        \delta(y - A_{b}\sin(qx - \omega t)) \text{d}x\text{d}t= \frac{L}{(A_{b}^{2} - y^{2})}
    \label{eq:density}
\end{align}
Combining Eqs.~\ref{eq:TA} and \ref{eq:density}, we obtain
\begin{align}
    T_A = \frac{\alpha}{2} \, v \,\xi_{\parallel} \, A_{b}^{2}L
    \label{eq:TA2}
\end{align}
This torque is balanced by the viscous rotational drag $\xi_{R,F}\Omega$ of the flagellum. 
For small beat amplitudes, the rotational drag coefficient of the flagellum is approximately 
equal to that of a rod of the same length, $\xi_{R,rod}=\xi_{\perp} L^{3}/12$. Moreover, 
if the polarity vector 
${\bf P}$ (swim direction) is not perpendicular to the gradient, but is inclined 
at an angle $\theta$, then it is the projection of mass distribution onto the axis 
perpendicular to the gradient direction $(\nabla\eta)/\eta_0$ which is relevant.
The balance of these two torques finally results in (see SI for details \cite{supp})
\begin{align} 
    \Omega 
    &=6 \, \frac{v}{L} \, \frac{\xi_{\parallel}}{\xi_{\perp}}\, 
         \left(\frac{A_b}{L}\right)^{2}\, \alpha\cos(\theta) \, \\    
        &= 6\, v\,\frac{\xi_{\parallel}}{\xi_{\perp}}\, 
    \left(\frac{A_b}{L}\right)^{2}\, 
    {\bf P} \times \nabla \left(\frac{\eta}{\eta_0}\right) \, .
    \label{eq:omegaanalytical}
\end{align}

The approximation of the rotational drag coefficient $\xi_{R}$ of the flagellum by
$\xi_{rod}$ of a stiff rod can be improved for larger beat amplitudes $A_b$ by a correction 
factor for the rotational drag coefficients $\xi_{R,rod}/\xi_{R,F}$, which is obtained directly 
from simulations. 
We calculate the rotational torque for a simulated configuration to obtain an angular velocity 
$\Omega$, and relate it to the rotational friction as
\begin{align}
 \xi_{R,F}=   T_R/\Omega = \left[\sum_{i} {\bf f}_{i} \times ({\bf r}_{i} - {\bf r}_{c})\right] /\Omega
\end{align}
Finally, this is averaged over one beat period, $\langle \xi_{R,F} \rangle_{\tau}$.
The rotational friction coefficient $\xi_{R,F}$ is smaller than $\xi_{R,rod}$, because the beating
flagellum is less extended, and some parts have orientational components parallel to the rotational
velocity. Moreover, for further analysis and comparison with simulation data,  we use a modified 
form of $\Omega$ which has an additional factor $\xi_{R,rod}/\xi_{R,F}$, i.e.,
\begin{align} 
    \Omega = 6 \frac{v}{L}\, \frac{\xi_{R,rod}}{\xi_{R,F}} \, \frac{\xi_{\parallel}}{\xi_{\perp}} \, 
    \left(\frac{A_b}{L}\right)^{2}\, \alpha\cos(\theta)
    \label{eq:omegaanalytical_corr}
\end{align}
This rotational velocity is influenced by three key factors: (i) the magnitude of the 
viscosity gradient $\alpha\cos(\theta) \propto\nabla \eta$, as already shown above to nicely agree 
with the simulations, (ii) the beat amplitude $A_b$, and (iii) via the velocity, 
on the wavelength $\lambda$.

\begin{figure}
    \centering
    \includegraphics[width=0.43\textwidth]{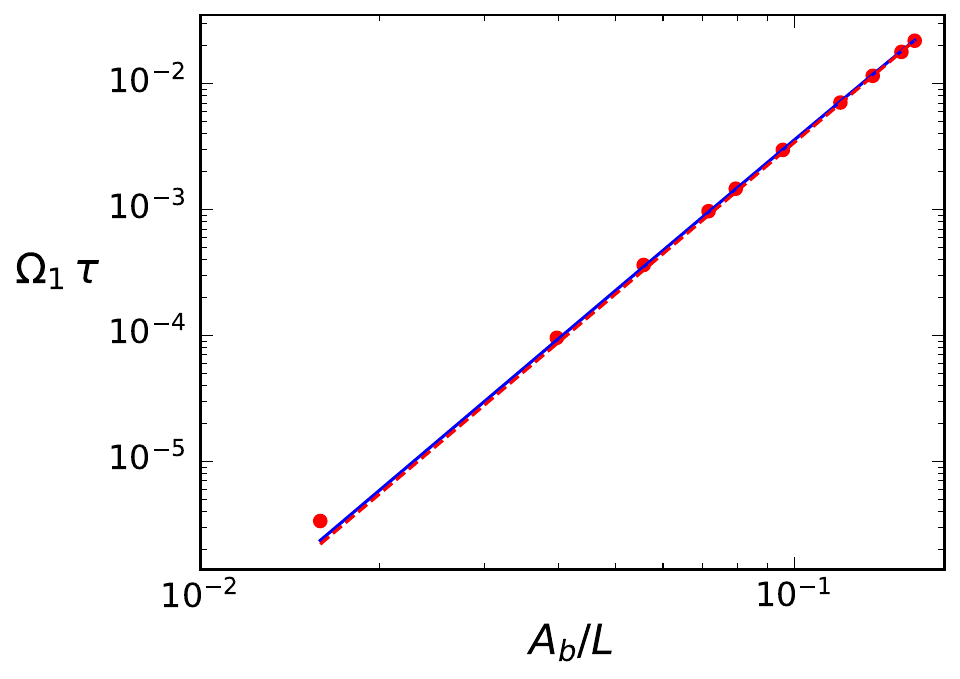}
    \caption{\textbf{Rotational velocity and beating amplitude relationship.} Rotational velocity $\Omega_1$ of the flagellum as a function of beating amplitude for 
    $\alpha = 0.4$. A fit (dashed red lined) of the simulation data (red dot symbols) to the functional 
    dependence $\Omega_1 \tau \sim (A_{b}/L)^4$ (see text)
    and the analytical predictions (blue solid line) of Eq.~\eqref{eq:powerlaw} are shown to agree very well.}
    \label{fig:omegaanalytical}
\end{figure}

In order to test the validity and accuracy of our analytical prediction \eqref{eq:omegaanalytical_corr} 
for $\Omega$, we compare it with simulation data. We consider first the dependence of 
$\Omega_1$ on the beat amplitude $A_{b}$, while keeping all other parameters constant.
Taking into account that the velocity depends quadratically on $A_b$  and inversely on wavelength $\lambda$,
see Eq.~\eqref{eq:GH}, the rotational velocity \eqref{eq:omegaanalytical_corr} is expected to scale as 
\begin{equation}
\Omega_1 \tau \sim \omega\tau \, (A_{b}/L)^{4} \, L/\lambda. 
\label{eq:powerlaw}
\end{equation}
Figure~\ref{fig:omegaanalytical} demonstrates that indeed $\Omega \sim A_b^4$. Thus, 
viscotaxis is predicted to display a very pronounced dependence on the beat amplitude.

\begin{figure}
    \begin{subfigure}{0.48\textwidth}
        \includegraphics[width=0.92\textwidth]{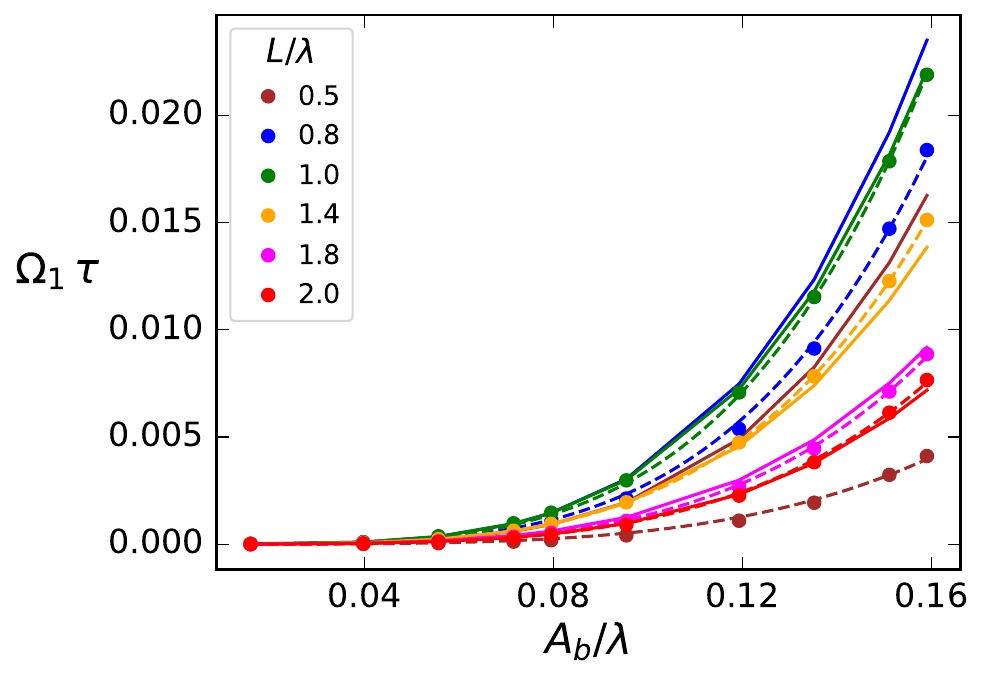}
     \end{subfigure}
    \hfill
    \hspace*{-0.01\textwidth}
    \begin{subfigure}{0.477\textwidth}
    \includegraphics[width=0.917\textwidth]{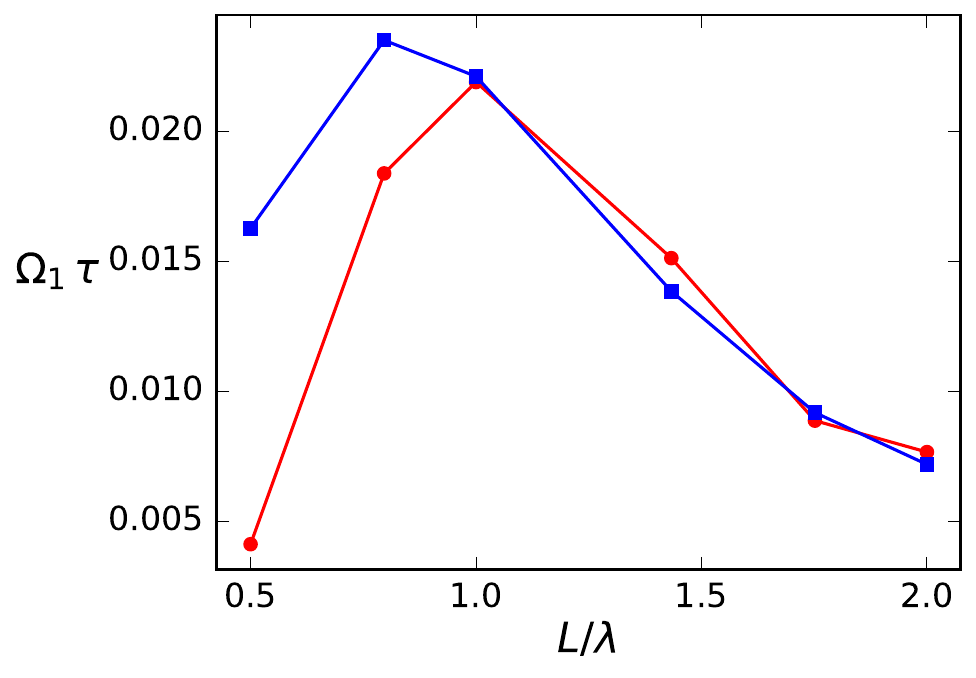}
     \end{subfigure}
\caption{\textbf{Rotational velocity as a function of beat amplitude and flagellum length.} 
    (a) Rotational velocity $\Omega_1 \tau$ as a function of normalised beat amplitude $A_{b}/\lambda$ 
    across different flagellum lengths $L/\lambda$ for simulation (dotted lines) and analytical results 
    (solid lines). The dashed lines are the fit to simulation data, with $\alpha  = 0.4$. 
    (b) Dependence of rotational velocity $\Omega_{1}\tau$ on flagellum length $L/\lambda$, for a fixed 
    wavelength $\lambda$ and normalized amplitude $A_b/\lambda = 0.16$, here red solid line with red 
    dot markers represents the simulation data and analytical result is shown by the blue solid line with 
    blue squares markers.}
    \label{fig:differentlength}
\end{figure}

Figure~\ref{fig:differentlength} illustrates the dependence of the reorientation rate $\Omega_1$ on the flagellar 
length $L$. Here, the flagellar length is varied for constant wavelength $\lambda$. For 
$L > \lambda$ ($L < \lambda$) there is more (less) than a single wave on the flagellum. 
The comparison in  Fig.~\ref{fig:differentlength}(a) demonstrates that there is in general good agreement 
between the analytical results \eqref{eq:omegaanalytical_corr} and simulation for $L>\lambda$. For $L<\lambda$, the observed deviations can be attributed to the assumptions inherent during the derivation of 
the analytical model; in this regime, the flagellar shape is no longer a travelling sine wave, but
rather a short rod-like filament oriented in the swimming direction, which bends left and right periodically. 
This dependence of $\Omega_1$ on $L/\lambda$ can be seen more clearly in Fig.~\ref{fig:differentlength}(b).
For $L/\lambda > 1$, the decrease of $\Omega_{1}$  with increasing $L$ is due to the 
increasing rotational drag $\xi_{R,F} \propto L^3$, as described by Eq.~\ref{eq:omegaanalytical_corr}. 
For $L/\lambda < 1$, swimming is impaired due to the absence of a travelling wave, as explained above, so that $\Omega_1 \to 0$ for $L/\lambda \to 0$.
This implies that the wave length for optimal viscotasis is $\lambda \simeq L$, as 
confirmed by the simulation results in Fig.~\ref{fig:differentlength}(b).

\subsection{Asymmetric Flagellar Beat}

\subsubsection{Viscotaxis for a Flagellum with Spontaneous Curvature}

We model asymmetric beat shapes by introducing a non-zero average spontaneous curvature $C_{0}$ in 
Eq.~\eqref{eq:curv}. This asymmetry reflects the experimental observation that sperm swim in circles or 
on helical trajectories \cite{gong2020steering} in the absence of a viscosity gradient. The spontaneous 
curvature generates beat patterns that break the symmetry 
and leads to a non-zero rotational velocity $\Omega_{0}$ of the flagellum in Eq.\eqref{eq:omegaofalpha} 
in the absence of a viscosity gradient, and consequentially to a circular trajectory of the center 
of mass with clockwise and anti-clockwise motion 
for $C_{0} > 0$ and $C_{0} < 0$, respectively \cite{gong2020steering}. Our analysis focuses on the 
scenario where the average spontaneous curvature $C_{0}$ is constant and does not vary in reaction to 
environmental conditions. 

\begin{figure}
    \centering
    \includegraphics[width=0.453\textwidth]{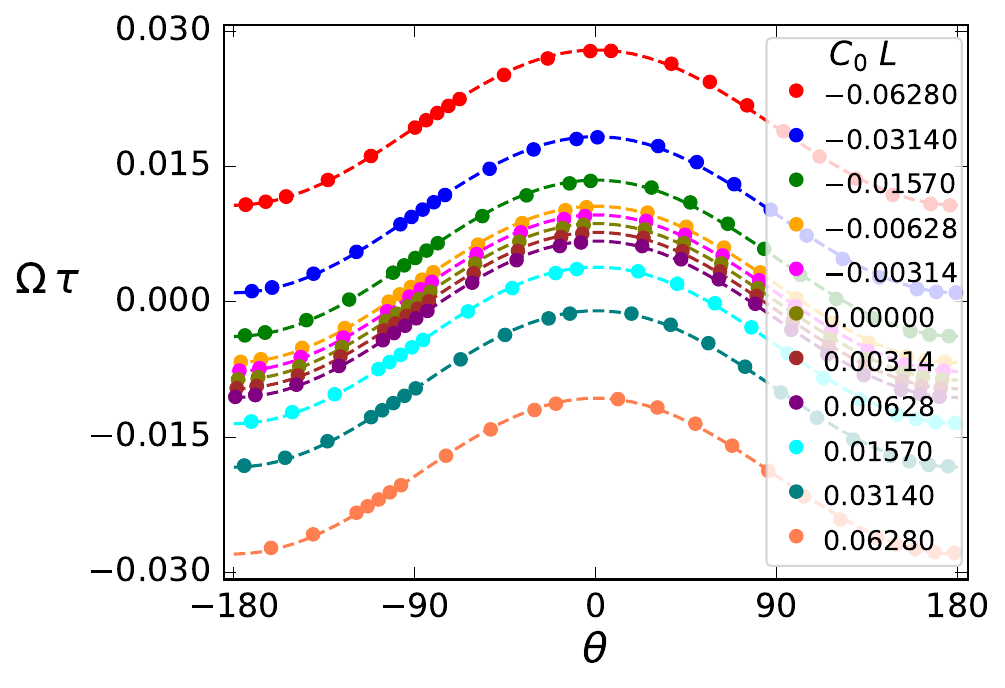}
    \caption{\textbf{Rotational velocity with spontaneous curvature.} Variation of $\Omega \tau$ as a function of $\theta$ for different values of spontaneous 
    curvature, with $\alpha = 0.4$. The peak value, $\Omega(0) \tau$, increases
    with decreasing $C_{0} L < 0$ and decreases with increasing $C_{0} L > 0$.}
    \label{fig:curvomega}
\end{figure}

We examine the effect of beat asymmetry on the viscotactic response.  The simulation results in 
Fig.~\ref{fig:curvomega} illustrate that the rotational velocity curve $\Omega(\theta)$ 
is shifted upwards or downwards for $C_{0} L<0$ or $C_{0} L>0$, respectively. 
Therefore, the total rotational velocity $\Omega$ can be described as a superposition 
of the contributions from $C_{0}$ and $\nabla\left(\frac{\eta}{\eta_0}\right)$,
\begin{equation}
    \Omega = \Omega_{0} + \Omega_{1} \alpha \cos(\theta) \, .
    \label{eq:twoomega}
\end{equation}

\subsubsection{Phase Diagram and Stability}

\begin{figure}
    \centering
    \includegraphics[width=0.41\textwidth]{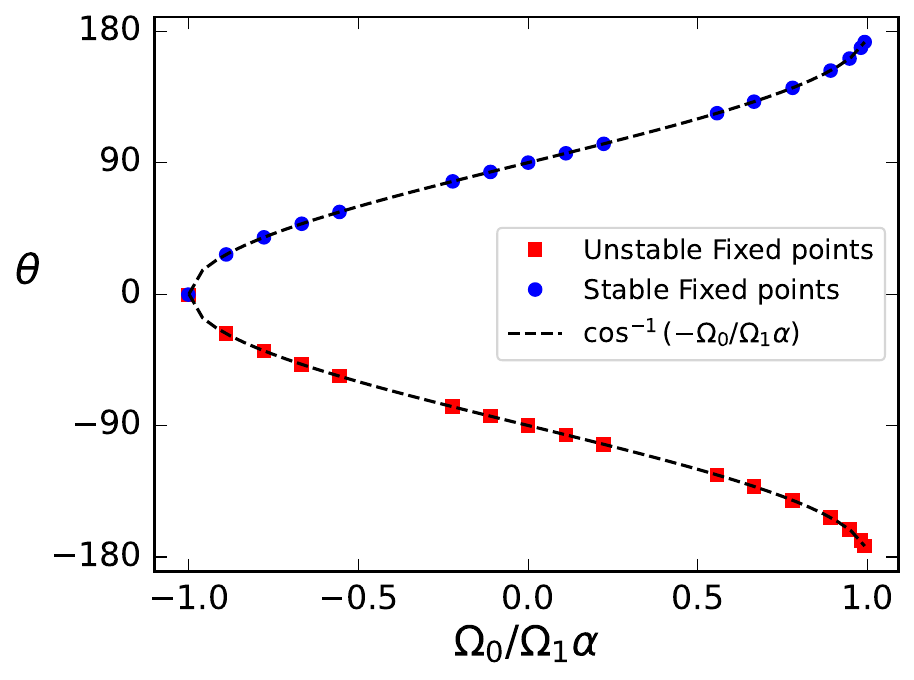}
    \caption{\textbf{Stability analysis of fixed points.}
        Variation of the fixed points ($\theta$) as a function of the ratio 
        ($\Omega_0 / \Omega_1\alpha$). Stable fixed points from simulation data are indicated by blue circles,
        while unstable fixed points are represented by red squares. The theoretical prediction given by 
        $(\cos^{-1}(-\Omega_0 / \Omega_1\alpha))$ is shown as a black dotted line. The stability of the fixed 
        points changes with $(\Omega_0 / \Omega_1\alpha)$, highlighting the transition between stable 
        and unstable regimes in the dynamics.}
    \label{fig:bifurcation}
\end{figure}

\begin{figure}
    \centering
    \includegraphics[width=0.453\textwidth]{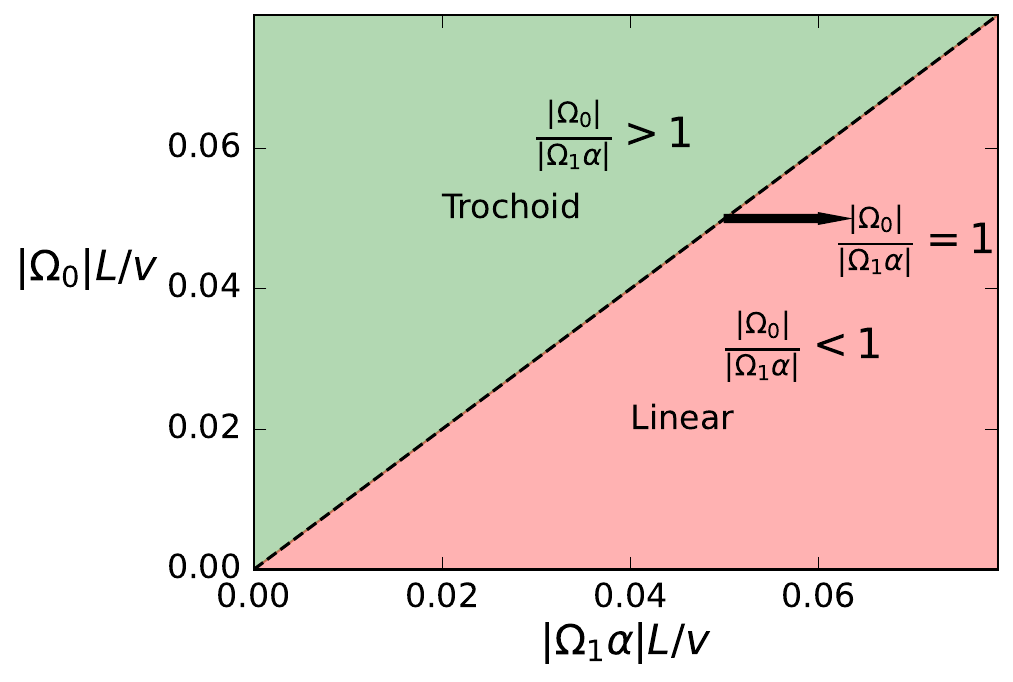}
    \caption{\textbf{Phase diagram of trajectory types.} Regions of linear and trochoid-like trajectories in a parameter 
    space defined by dimensionless quantities $|\Omega_{0}|L/v$ and $|\Omega_{1}\alpha| L/v$. The demarcation 
    line ${|\Omega_{0}|}/{|\Omega_{1}\alpha|}=1$ as indicated by the black dashed line across the plot 
    separates the two distinct regimes. For ${|\Omega_{0}|}/{|\Omega_{1}\alpha|}>1$ the system exhibits 
    trochoid-like motion, while for ${|\Omega_{0}|}/{|\Omega_{1}\alpha|}<1$ the system remains stationary.}
    \label{fig:phasediagram}
\end{figure}

The stability analysis of the fixed points of Eq.~\eqref{eq:twoomega} reveals a double saddle-node 
bifurcation, as shown in Fig.~\ref{fig:bifurcation}. 
For $|\Omega_{0}|<|\Omega_{1} \alpha|$, there are a stable and an unstable fixed point $\theta^*$ 
with $\Omega(\theta^*) = 0$, so that
\begin{align}
    \cos(\theta^{*}) = -\frac{\Omega_{0}}{\Omega_{1} \alpha} 
\end{align}
at which the flagellar trajectory is asymptotically a straight line. 
For $\Omega_0=0$, the flagellum moves up the gradient at the stable fixed point with $\theta=\pi/2$, 
while it moves down the gradient at the unstable fixed point with $\theta=-\pi/2$.
As spontaneous curvature and $\Omega_0$ decreases ($\Omega_{0}/(\Omega_{1}\alpha) \to -1$), both fixed 
points move towards $\theta=0$ (bifurcation point).
At $\Omega_{0}=-\Omega_{1} \alpha$, the two fixed points meet, resulting in a straight trajectory 
perpendicular to and to the right of the gradient direction to the right. 
The second bifurcation happens accordingly in the opposite direction for 
$\Omega_{0}/(\Omega_{1} \alpha) \to +1$.
For $|\Omega_{0}|>|\Omega_{1} \alpha|$, no fixed point exist, and the flagellum constantly rotates,
as discussed below.
With this classification and numerical results, we can construct the phase diagram 
displayed in Fig.~\ref{fig:phasediagram}. 

\begin{figure}
    \begin{subfigure}{0.45\textwidth}
         \includegraphics[width=0.45\textwidth]{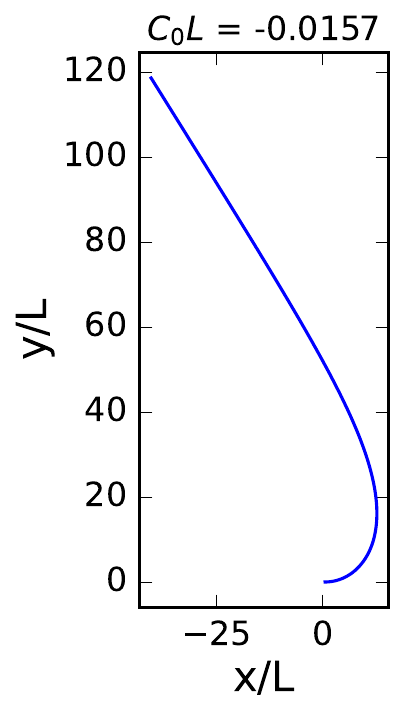}
     \end{subfigure}
    \begin{subfigure}{0.45\textwidth}
         \includegraphics[width=0.9\columnwidth]{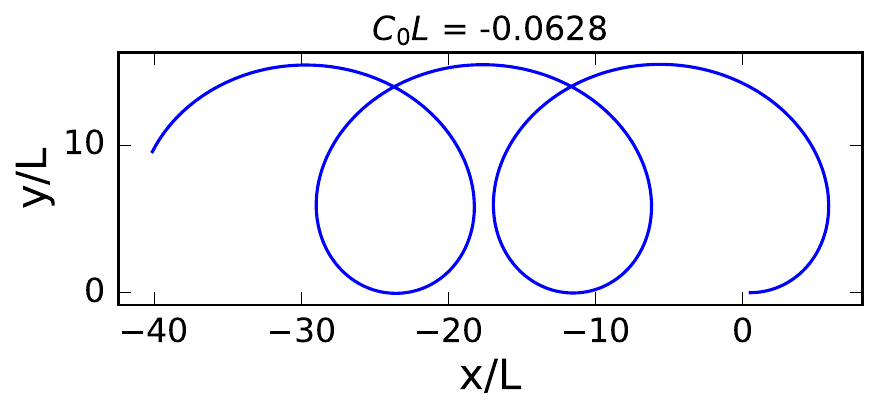}
     \end{subfigure}
\caption{\textbf{Trajectory of flagella with spontaneous curvature.} Trajectory for the center of mass of beating 
    flagella with average spontaneous curvature (a) $C_{0}L = -0.0157$ with a stable fixed point, moving on an 
    asymptotically straight trajectory, and (b) $C_{0}L = -0.0628$ with no fixed point, showing period motion on 
    a trochoid-like trajectory.  See also movies M2 and M3 \cite{supp}.}
\label{fig:differenttrajectory}
\end{figure}

To illustrate this behavior, we select two curves from Fig.~\ref{fig:curvomega}, with two spontaneous 
curvatures $C_{0}L = -0.0157$ with $\Omega_{0}<\Omega_{1} \alpha$ and two fixed points, and 
$C_{0}L = -0.0628$ with $\Omega_{0}>\Omega_{1}\alpha$) and no fixed point. The corresponding 
trajectories are shown in Fig.~\ref{fig:differenttrajectory}. At the stable fixed point, 
$\cos(\theta^{*}) = -\Omega_{0}/\Omega_{1} \alpha$, the two curvature mechanisms oppose 
each other, and the flagellum moves on an asymptotically straight trajectory with inclination 
$\theta^{*}$. For large $\Omega_0$, the spontaneous curvature dominates, and a 
trochoid-like trajectory 
emerges, with drifting circular motion perpendicular to the gradient.

\subsubsection{Self-Propelled Particle Model and Perpendicular Drift}

To characterize the trochoid-like drift motion in the phase diagram of Fig.~\ref{fig:phasediagram} 
in more detail, we analyze the pitch and the drift direction, which are affected by the 
viscosity gradient $\alpha$ and spontaneous curvature $C_{0}$. 

The dynamics described by Eq.~\eqref{eq:twoomega} is the 
motion of a self-propelled particle (SPP) in the non-thermal limit, 
with constant speed $v$ and a propulsion direction ${\bf e}(t)$, which in polar 
coordinates is ${\bf e}=(\cos(\theta), \sin(\theta))$ with inclination angle $\theta$. 
Under the effect of a redirectional torque, the orientational motion of this particle is 
governed by
\begin{align}
    \dot{\theta} = \Omega_{0} + \Omega_{1} \alpha \cos(\theta)
    \label{eq:abptheta}
\end{align}
The particle position ${\bf r}(t)$ is then obtained from the equation of motion
\begin{align}
    \dot{{\bf r}}(t) = v\,\boldsymbol{e}(t)
    \label{eq:apbposition}
\end{align}

The trochoid-like motion is a  periodic motion, compare Fig.~\ref{fig:differenttrajectory}(b), with time 
period $\Delta T$. The pitch, denoted as $P$, is the repeat distance along the drift direction, 
calculated over one complete cycle or one time period, from $\theta= 0$ to $2\pi$, i.e., 
\begin{align}
    P = |\Delta {\bf r}(t)| =  | \int_{0}^{T}\dot{\boldsymbol{r}}(t) dt | 
\end{align}
The time period $\Delta T$ of each cycle can be calculated from Eq.~\eqref{eq:abptheta} as
\begin{equation}
    \Delta T = \int_{0}^{2\pi} d\theta \frac{1}{\Omega_{0} + \Omega_{1} \alpha \cos(\theta)} 
        = \frac{2\pi}{\sqrt{(\Omega_{0})^{2} - (\Omega_{1}\alpha)^{2})}}
    \label{eq:timeperiod}
\end{equation}
where $|\Omega_{0}| > |\Omega_{1}\alpha|$. The explicit dependence of $t(\theta)$ on $\theta$ is 
obtained from the indefinite integral of Eq.~\ref{eq:abptheta} (see SI for detail \cite{supp}).
This shows that increasing $\Omega_{0}$ reduces the cycle time, while increasing $\Omega_{1} \alpha$ 
extends the cycle time. The cycle time diverges as the fixed point is approached.

To quantify the pitch, we employ perturbation theory, with expansion parameter 
$\epsilon = \Omega_{1}\alpha/\Omega_{0}$, which measures the deviation from purely circular motion.
Thus, we study the equation
\begin{align}
    \frac{d\theta}{dt} = \Omega_{0}(1 + \epsilon \cos(\theta)) \, .
    \label{eq:epsilon}
\end{align}
by employing the ansatz $\theta(t) = \theta_0(t) + \epsilon \theta_1(t)$.
For $\epsilon = 0$, the solution of Eq.~\eqref{eq:epsilon} is $\theta_{0}(t) = \Omega_0 t$, describing a 
circular trajectory. The leading-order contribution $\theta_1$ for small $\epsilon$ is then determined by
\begin{equation}
\frac{d\theta_{1}}{dt} = \Omega_{0} \cos(\Omega_{0}t)
\end{equation}
The solution $\theta_{1}(t) = \sin(\Omega_0 t)$ is used to calculate the translational 
drift
\begin{align}
    \Delta x &= \int_{0}^{t^{'}} v \cos(\theta_{0} + \epsilon\theta_{1}) dt = -v \pi\epsilon/|\Omega_{0}|
    \label{eq:pitchx}\\
    \Delta y &= \int_{0}^{t^{'}} v \sin(\theta_{0} + \epsilon\theta_{1}) dt = 0 
    \label{eq:pitchy}
\end{align}
where, $\epsilon = \Omega_{1}\alpha/\Omega_{0}$ and $t^{'} = 2\pi/(|\Omega_{0}|(\sqrt{(1 - \epsilon^{2})})$.
This results demonstrates two important features of the trochoid-like motion:
(i) there is no drift along the $y$-axis, which shows that the particle always drifts 
{\em perpendicular} to the gradient direction, and (ii) the resulting pitch is $P = -v\pi\epsilon/|\Omega_{0}|$. 

\begin{figure}
    \centering
    \includegraphics[width=0.442\textwidth]{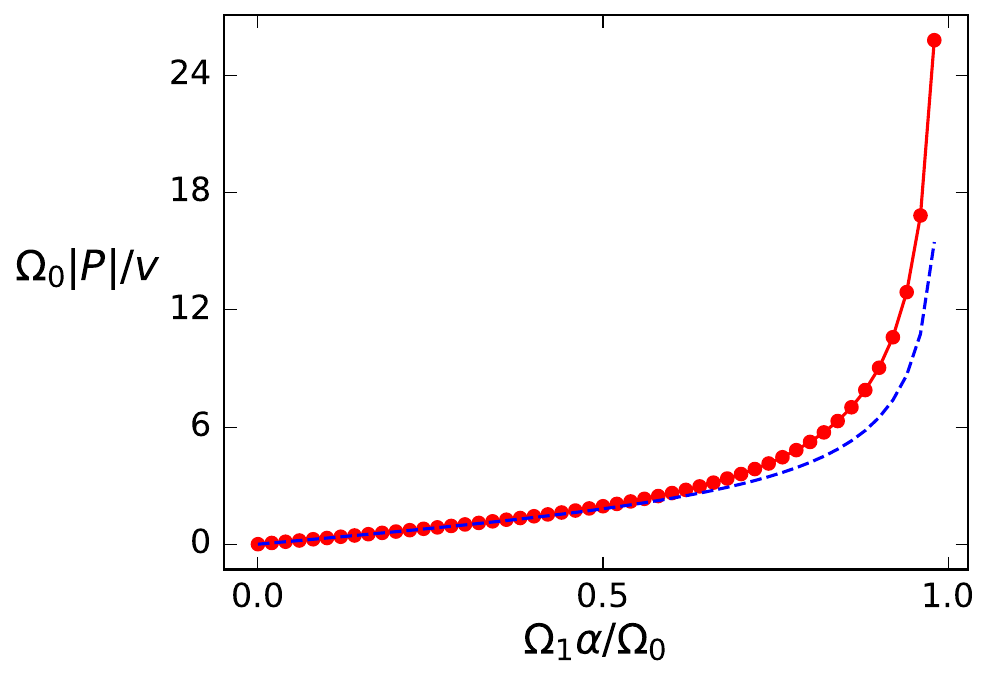}
    \caption{{\bf Pitch of trochoid-Like drift motion.} Comparison between theoretical prediction and simulation results for the pitch $\Omega_{0}|P|/v$ 
    as a function of the ratio $\Omega_{1}\alpha/\Omega_{0}$. The dotted red line with markers represents 
    the simulation data, the dashed blue line the theoretical predictions. Good agreement 
    is obtained over the whole accessible range of $\Omega_{1}\alpha/\Omega_{0}$. }
    \label{fig:pitch}
\end{figure}

We can generalize this solution for the pitch by combining the first-order perturbation 
theory with the observation that the pitch must diverge for $\Omega_{1}\alpha  \to \Omega_{0}$,
and vanishes for $\Omega_{1}\alpha = 0$, $\Omega_{0} \neq 0$, which results in the
approximation
\begin{align}
    P = \frac{-v\Omega_{1} \alpha \pi}
              {\Omega_{0}|\Omega_{0}|\sqrt{1 - (\Omega_{1} \alpha/\Omega_{0})^{2}}}
\end{align}
To investigate the quality of this analytical approximation, we compare it with the simulation results, 
as shown in Fig.~\ref{fig:pitch}. The agreement is quite satisfactory over the whole accessible range of 
$\Omega_1\alpha/\Omega_0$ values.

\subsection{Effect of Flagellar Elasticity on Viscotaxis}
\label{sec:elasticity}

A flagellum is not a rigid object, but has a finite elasticity and deformability.
The elasto-dynamics of a flexible filament can be conveniently characterized by a single 
dimensionless parameter, the sperm number $Sp$ \cite{moreau2018asymptotic, kumar2019effect, gauger2006numerical, lowe2003dynamics, alizadehrad2015simulating,yu2006experimental,lauga2007floppy}, 
\begin{align}
    Sp^{4}=T_{v}/\tau = L^{4} \xi_{\perp} \omega/\kappa ,
    \label{eq:spermnumber}
\end{align}
which can be interpreted as the ratio of two-time scales, the beat period $\tau =2\pi/\omega$ 
and the typical relaxation time $T_{v} = 2 \pi L^{4} \xi_{\perp}/\kappa$ of an elastic 
perturbation to propagate along the whole flagellar length $L$ in a viscous environment.

In order to investigate the dependence of the rotational velocity on the activity-induced deformations 
of the flagellum, we study the relation between the rotational velocity $\Omega_1$ and the sperm number $Sp$
by variation of the beat frequency $\omega$ for several combinations of bending rigidity $\kappa$ 
and perpendicular friction coefficient $\xi_\perp$.  
In the limit of $Sp \rightarrow 0$, rigidity dominates, and the shape of the flagellum follows exactly 
the prescribed curvature pattern, whereas at large $Sp$, viscosity dominates, resulting in additional 
deformations of the filament.

These additional deformations of filament change the effective beating amplitude of the flagellum. 
In order to predict this dependence, we employ the curvature energy for the flexible worm-like filament 
as a function of its conformation \cite{doi_1986_book},
\begin{align}
    \mathcal{H} = \frac{\kappa}{2}\int_{0}^L(\partial_x^{2}h(x,t) - C_{0}^{'}(x,t)) dx
    \label{eq:helfrich}
\end{align}
for a nearly straight flagellum parallel to the $x$-axis.
Here, $h(x,t)$ represents small perpendicular excursions at a distance $x$ along the filament. 
$C_{0}^{'}(x,t)$ denotes 
the local spontaneous curvature, where a non-zero value means the filament would spontaneously relax 
to $C_{0}^{'}$, which is a minimum-energy state. The local spontaneous curvature 
has the same role as described in Eq.~\eqref{eq:curv}. We ignore contributions from boundary 
terms, as they do not influence the subsequent physics.
The dynamics of the filament is then governed by the local viscous drag and bending force, derived 
from Eq.~\eqref{eq:helfrich}, and is described by the Langevin equation
\begin{align}
    \xi_{\perp}\frac{\partial h(x,t)}{\partial t} = - \frac{\delta \mathcal{H}}{\delta h(x,t)} 
               = -\kappa(\partial_x^{4}h - \partial_x^{2}{C_{0}^{'}})
    \label{eq:langrange}
\end{align}

\begin{figure}
    \centering
    \includegraphics[width=0.438\textwidth]{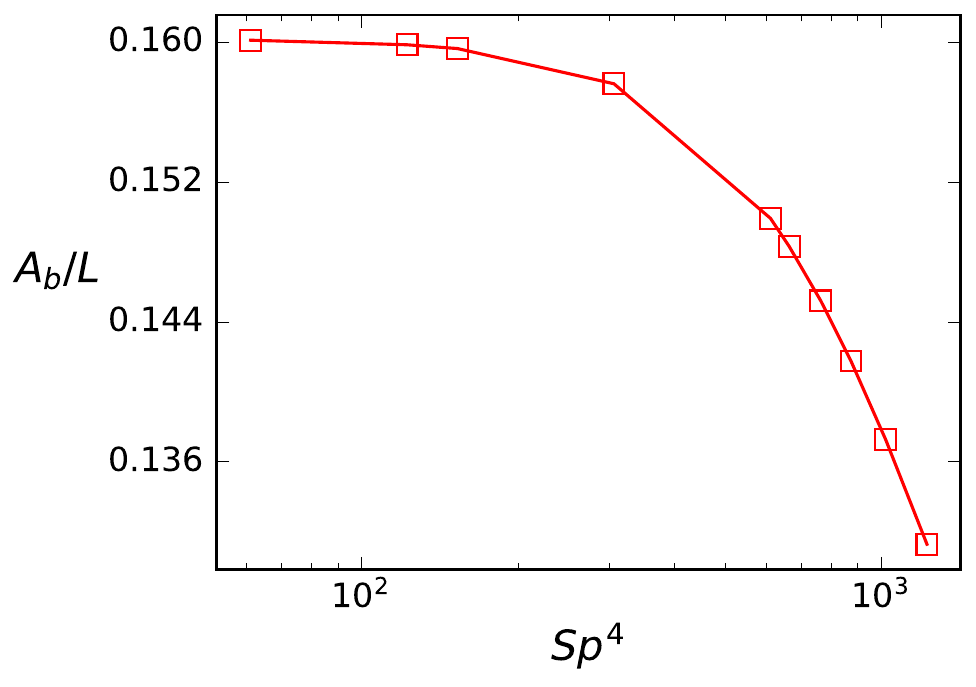}
    \caption{\textbf{Deependence of beating amplitude on sperm number.} Beating amplitude $A_{b}$ of the 
    flagellum as a function of $Sp^4$. Data for various beat 
    frequencies $\omega$, perpendicular friction coefficients $\xi_\perp$, and bending rigidities 
    $\kappa$ all fall onto a single master curve.}
    \label{fig:amplitudesp}.
\end{figure}

To solve Eq.~\eqref{eq:langrange} with a travelling spontaneous-curvature wave, 
$C_{0}^{'}(x,t) = \bar{C_{0}}e^{i(qx-\omega t)}$, we use a travelling-wave ansatz also for the 
deflection, $h(x,t) = h_{0}e^{i(qx-\omega t)}$, where $h_{0} = |h_0| e^{i\phi}$ represents 
the (complex) beating amplitude. This implies 
\begin{align}
    \xi_{\perp} i \omega h_{0} & = \kappa(q^{4}h_{0} + q^{2}\bar{C_{0}})\\
    h_{0} & = \frac{q^{2}\bar{c_{0}}L^{4}}{(iSp^{4} - q^{4}L^{4})}
    \label{eq:ampli_complex}
\end{align}
Therefore, the beating amplitude for the beating flagellum is
\begin{align}
    \frac{|h_{0}|}{L} = \frac{(qL)^{2}(\bar C_{0}L)}{((qL)^{8} + Sp^{8})^{\frac{1}{2}}}
    \label{eq:beatampl_sp}
\end{align}
Equation~\eqref{eq:beatampl_sp} shows that the beating amplitude increases linearly with 
$\bar{C_{0}L}$ and decreases monotonically as a function of $Sp^{4}$. In the limit of small sperm 
number ($Sp \to 0 $), $|h_{0}|/L = \pi^{2} \lambda \bar{C_{0}} L/4$. 
A strong decrease of the beat amplitude with increasing $Sp$, as predicted by 
Eq.~\eqref{eq:beatampl_sp}, is indeed observed in the simulations, see Fig.~\ref{fig:amplitudesp}.
From Eq.~\eqref{eq:ampli_complex}, we can also extract the phase-lag $\phi$ of the beat 
wave relative to the spontaneous curvature wave, 
\begin{align}
    \sin(\phi) = \frac{Sp^{4}}{((qL)^{8}+Sp^{8})^{\frac{1}{2}}}
\end{align}
This implies that $\phi = 0$ for $Sp \to 0$, i.e. the beat and spontaneous-curvature waves are
in phase, while $\phi = \frac{\pi}{2}$ for $Sp \to \infty$, i.e. the two waves have the 
maximum phase shift.

\begin{figure}
    \begin{subfigure}{\columnwidth}
         \includegraphics[width=0.89\columnwidth]{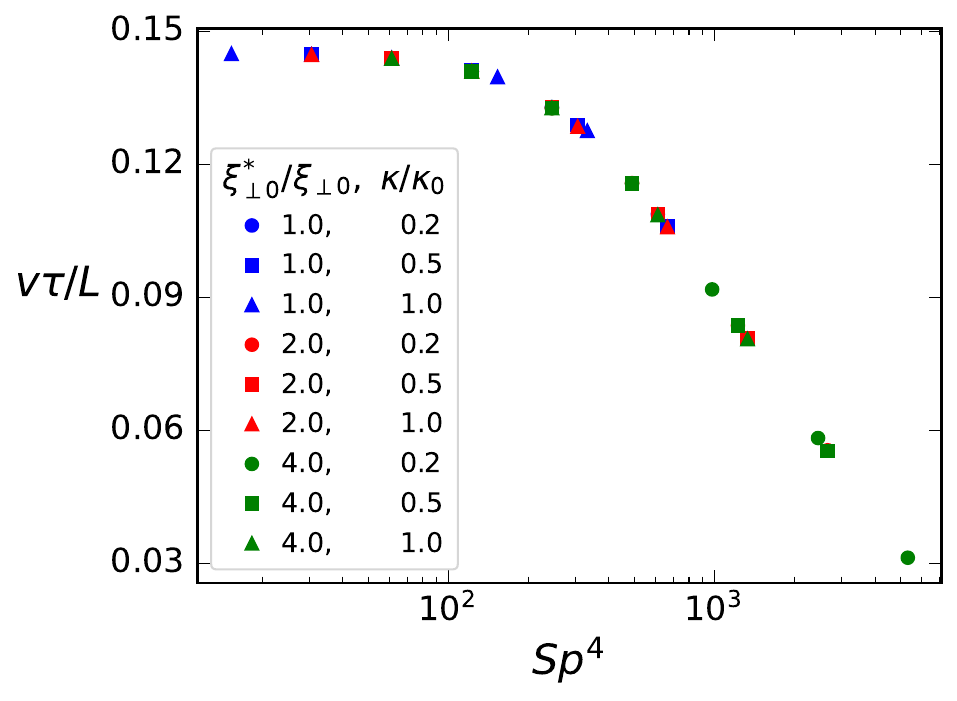}
     \end{subfigure}
    \begin{subfigure}{\columnwidth}
         \includegraphics[width=0.89\columnwidth]{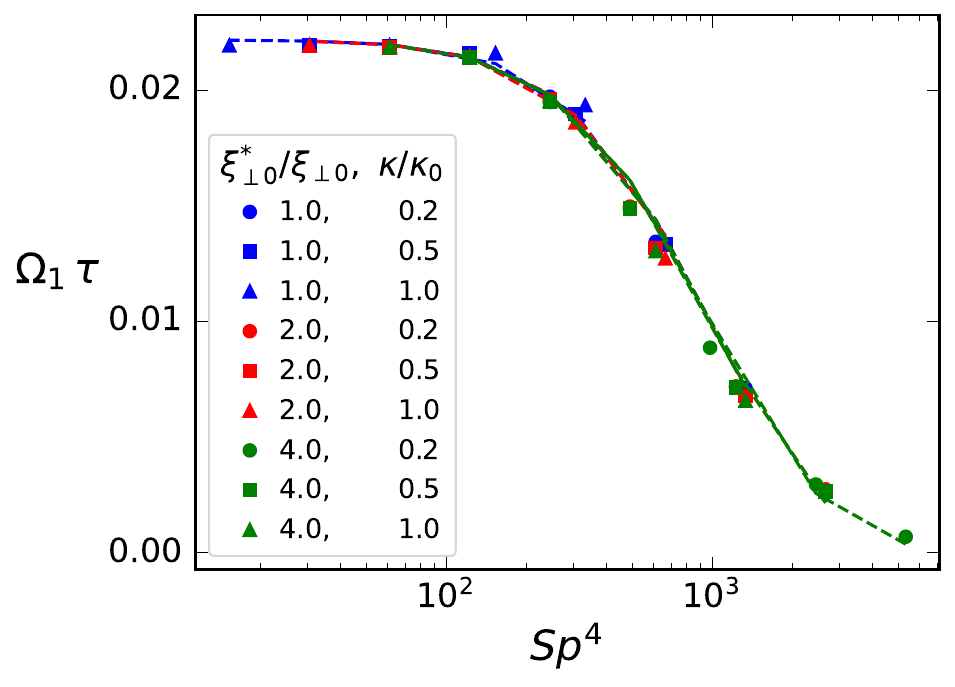}
     \end{subfigure}
\caption{(a) \textbf{Universal master curves for dependence of propulsion velocity and viscotatic response 
    on flagellar deformability.} 
    (a) Dimensionless propulsion velocity $v\tau/L$ of a flagellum as a function of $Sp^{4}$ for 
    different set of parameters ($\xi_{\perp}, \kappa$) and $\alpha = 0.0$. Symbols are simulation data. 
    (b) Normalized rotational velocity as a function of $Sp^{4}$ (with variation of the beat 
    frequency). The symbols are the simulation data, and dashed lines are the analytical prediction 
    for $\Omega_{1}$. As the sperm number decreases by decreasing beat frequency, the rotational velocity 
    approaches a plateau.}
    \label{fig:spermnumber}.
\end{figure}

Furthermore, this analytical dependence of beating amplitude on $Sp^{4}$, Eq.~\eqref{eq:beatampl_sp}, 
can be combined with the swimming velocity as a function of $Sp^{4}$, obtained numerically 
for $\alpha=0.0$ (Fig.~\ref{fig:spermnumber}(a)), to theoretically predict (by employing 
Eq.~\eqref{eq:omegaanalytical_corr}) the viscotactic reorientation rate $\Omega_1$ as a function of $Sp^{4}$.
The simulation results in Fig.~\ref{fig:spermnumber}(b) show that $\Omega_1 \tau$ for 
various beat frequencies $\omega$ and friction coefficients $\xi_\perp$ all fall onto a single master 
curve, which is described by a universal scaling function $\Gamma$, with 
\begin{equation}
\Omega_1\tau = \Gamma(Sp^4) \ .
\end{equation}
Moreover, the simulation results closely match the analytical results, see Fig.~\ref{fig:spermnumber}(b). 
In addition, $\Omega_1\tau$ is found to approach a finite value in the limit $Sp \to 0$, achieved by 
decreasing the beat frequency $\omega \to 0$. This is a non-trivial limit, because the rotation frequency 
vanishes, while the beat period diverges. The fact that the product $\Omega\tau$ remains finite indicates 
that the rotation of the flagella does not require a dynamic deformation of the flagellum, i.e. the 
rotational motion is not a consequence of changes in the shape during its beating motion, but due to the 
friction asymmetry discussed in Sec.~\ref{sec:barrierssph} above, see Eq.~\eqref{eq:omegaanalytical}. 

\begin{figure}
    \centering
    \includegraphics[width=0.45\textwidth]{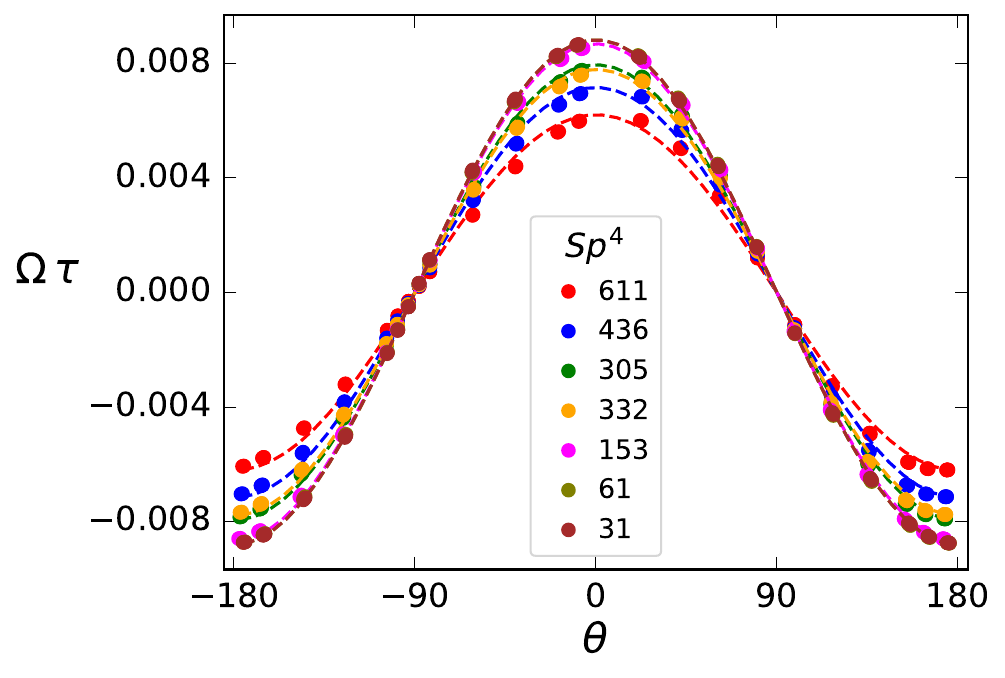}
    \caption{\textbf{Rotational velocity vs. orientation (\boldsymbol{$Sp^4$}).}
    Rotational velocity $\Omega \tau$ as a function of orientation $\theta$ at different $Sp^4$ for 
    $C_{0}L = 0$ and $\alpha= 0.4$.}
    \label{fig:spermnumberomega}
\end{figure}

The reduced viscotatic response of the flagellum due to elasticity and deformability  
is mainly due to a decreasing beat amplitude, which leads to a decreasing 
friction asymmetry and also a reduced propulsion speed (Gray and Hancock estimate $v/\omega \sim A_b^2$).
The combined effect on the viscotactic response is very pronounced due to the strong 
dependence $\Omega_1 \sim A_{b}^4$ on the beat amplitude, compare Eq.~\eqref{eq:powerlaw}. 
This effect also manifests itself in the angular dependence of the rotation rate $\Omega(\theta)$ in 
Fig.~\ref{fig:spermnumberomega}, where the magnitude of $\Omega$ decreases both at the minima and maxima --
in contrast to the case of spontaneous curvature, where the whole $\Omega(\theta)$ curve shifts upwards.
This implies that for elastic flagella, motion perpendicular to the gradient in either direction implies 
a slower reorientation in the gradient direction due to deformability.

\begin{figure}
    \centering
    \includegraphics[width=0.419\textwidth]{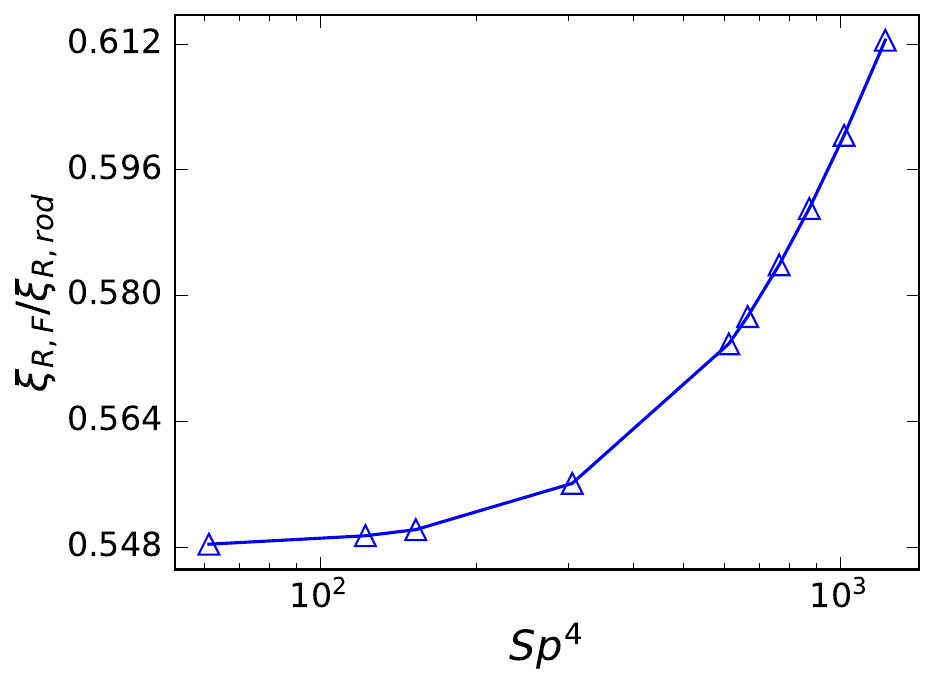}
    \caption{\textbf{Rotational friction coefficient vs. \boldsymbol{$Sp^4$}.} Variation of the ratio of
    rotation friction coefficients, $\xi_{R,F}/\xi_{R,rod}$, as a function of sperm
    number $Sp^4$. The coefficient increases with increasing $Sp^4$.}
    \label{fig:rotationalfrictionsp}
\end{figure}

The deformability of the flagellum also affects rotational friction coefficient $\xi_R$. 
The results in Fig.~\ref{fig:rotationalfrictionsp} show that $\xi_{R,F}$ increases with increasing flexibility. 
This contributes to the decrease of $\Omega_1$, because $\Omega_1 \sim 1/\xi_{R,F}$, compare 
Eq.~\eqref{eq:omegaanalytical_corr}.
In analogy with the rotational friction of a stiff rod, $\xi_{R,rod} \sim L^3$, we can attribute the increase of 
$\xi_{R,F}$ to a increment of the projected length, i.e. $\xi_{R,F} \sim L_{eff}^{3}$, where $L_{eff}$ is 
end-to-end length of the flagellum due to its deformation. Of course, the increases of the projected length of the flagellum and the decrease of the beat amplitude are geometrically related for fixed contour length of the 
flagellum, with $L-L_{eff} = \pi^2 (A_b/L)^2$ for $\lambda=L$ to leading order in $A_b/L$ (see SI for details \cite{supp}).

\section{Discussion and Conclusions}
\label{sec:conclusions}
We have studied the viscotactic behavior of flagella by simulating semi-flexible filament with an active
bending wave in two spatial dimensions. Our theoretical and simulation study focuses on the motion of the 
flagellum in an environment with a linear viscosity gradient. The flagellum is found to reorient toward regions 
of higher viscosity, where the rotational velocity increases for steeper gradients. We derive an approximate 
analytical expression for the reorientation rate, which matches our simulation data very well. Additionally,  
flagella with asymmetric beat determined by an average spontaneous curvature exhibit enhanced 
viscotaxis. For strong spontaneous curvature, positive viscotaxis is replaced by a trochoid-like drift
motion {\em perpendicular} to the gradient direction. Flagellar deformability, characterized by the 
dimensionless  sperm number $Sp$, is found to reduce the beat amplitude, swimming speed, and the viscotactic 
response. 

It is important to note that the viscosity gradient and the local
viscosity is kept constant in our study. This is of course not possible in a real system, when
the flagellum shows positive viscotaxis. In a real system, with constant viscosity gradient $\partial_y \eta$ and initial conditions corresponding to the ``positive viscotaxis" region of the phase diagram of Fig.~\ref{fig:phasediagram}, the flagellum moves to regions of higher viscosity, which implies that
$\alpha= L(\partial_y \eta)/\eta$ in Eq.~\eqref{eq:omegaanalytical} 
decreases along the trajectory, until it reaches $\theta=0$ (or $\theta=\pi)$ and moves perpendicular to the gradient. This happens when the separatrix in Fig.~\ref{fig:phasediagram} -- equivalent to the condition $\alpha=|\Omega_0/\Omega_1|$ -- is reached. This uniquely determines the asymptotic $y^*$-position given implicitly through 
\begin{equation}
\eta(y^*)= L (\partial_y \eta) \, |\Omega_1/\Omega_0|.
\end{equation}
This implies, in particular that for deep penetration into a region of higher viscosity, it is advantageous to have a small $\Omega_0$, i.e. a small spontaneous curvature and an nearly symmetrical beat. 

This study can be extended in several directions. First, sperm has not only a flagellum but
also a head, which affects the motion in a viscosity gradient. Second, sperm do move in the 
vicinity of surfaces, which is one of the motivation for our modelling in two spatial dimensions.
However, it would of course be very interesting to see and investigate the motion in linear
viscosity gradients in three dimensions. Finally, it would be important to study sperm
motion in more complex viscosity landscapes.

\section*{Author Contributions}
G.G. and J.E. designed the research project. S.A. performed the simulations and analysed the obtained
data. All authors performed analytic calculations, participated in the discussions and writing of the manuscript.

\section*{Conflicts of interest}
There are no conflicts to declare.

\begin{acknowledgments}
We gratefully acknowledge funding from the ETN PHYMOT (``Physics of Microbial Motility”) within
the European Union’s Horizon 2020 research and innovation programme under the 
Marie Sklodowska-Curie grant agreement No 955910, and computing time resources provided by 
Forschungszentrum J\"ulich.

\end{acknowledgments}





%

\end{document}